\documentstyle[aps,twocolumn,prl,epsf]{revtex}

\begin{document}

\newcommand{\be}{\begin{equation}}
\newcommand{\ee}{\end{equation}}

\draft

\twocolumn[\hsize\textwidth\columnwidth\hsize\csname @twocolumnfalse\endcsname

\title{Orbital Correlations in the Ferromagnetic Half-Metal $CrO_{2}$
}
\author{M. S. Laad$^1$, L. Craco$^2$ and E. M\"uller-Hartmann$^1$}

\address{${}^1$Institut f\"ur Theoretische Physik, Universit\"at zu K\"oln, 
Z\"ulpicher Strasse, 50937 K\"oln, Germany \\
${}^2$Instituto de Fisica Gleb Wataghin - UNICAMP, C.P. 6165, 
13083-970 Campinas - SP, Brazil   
}
\date{\today}
\maketitle

\widetext

\begin{abstract}
  We deduce a model relevant for the description of the 
ferromagnetic half-metal Chromium dioxide ($CrO_{2}$), widely used in 
magnetic recording technology.  The model describes the effect of dynamical,
local orbital correlations arising from local quantum chemistry of the 
material.  A finite temperature solution of the model in $d=\infty$ provides
a natural explanation of the optical response, photoemission, resistivity and 
the large Woods-Saxon ratio observed in experiments.  Our study confirms the
important role of many body dynamical correlation effects for a proper 
understanding of the metallic phase of $CrO_{2}$.
\end{abstract}

]

\narrowtext

  Colossal Magnetoresistive (CMR) materials have been the focus of renewed
theoretical and experimental investigations~\cite{[1],[2]} in recent years, 
given their obvious technological potential. Interest has mainly been centered 
around the manganites, where the interesting ferromagnetic metallic regime is 
driven by the double-exchange (DE) mechanism, though a unified understanding 
of the correlated nature of the metallic state itself, as well as the high-$T$ 
insulating paramagnet, requires consideration of orbital correlations and the
Jahn-Teller (JT) distortion on an equal footing with the DE ferromagnetism.

  Some attention has also focused on $CrO_{2}$, widely used in magnetic 
recording tapes.  In contrast to the manganites, stoichiometric $CrO_{2}$ 
is already a ferromagnetic metal.  Given the formal $4+$ valence state of 
$Cr$, the two $3d$ electrons occupy the $t_{2g}$ orbitals. One would 
intuitively expect to form $S=1$ spin on each site, and an antiferromagnetic 
Mott insulator.  Why $CrO_{2}$ is a ferromagnetic metal instead, has been 
answered by Korotin {\it et al}~\cite{[3]}, who have carried out insightful 
(LDA + U) calculations for this material.  Their main conclusions are: (i) 
the $O$ $2p$ band(s) act, atleast partially, as electron (or hole) reservoirs 
resulting in $Cr$ being mixed-valent (like $Mn$ in doped manganites), 
explaining metallicity, (ii) an almost dispersionless majority spin band of 
predominantly $d$ character at about $1 eV$ below $E_{F}$ over a large region 
of the Brillouin zone.  This corresponds to strongly localized $xy$ orbitals 
completely occupied by one majority spin electron.  On the other hand, the 
$d$ states of predominantly $d_{yz+zx}$ character hybridize with the $O$ $2p$ 
band and disperse, crossing $E_{F}$.  The Hund's rule coupling between the 
localized $d_{xy}$ spin and the spin density of the band $d_{yz+zx}$ electrons 
polarizes the latter, giving a ferromagnetic state via the double-exchange 
mechanism.  Thus, both the metallicity and ferromagnetism are correlated well 
with each other.

  A closer examination reveals that the metallic state of $CrO_{2}$ is 
strongly correlated, implying that many-body correlation effects beyond the
local-density approximation (LDA)~\cite{[4]} (or its variants) need to be 
considered. A number of experimental observations tend to support such a 
picture:

(1) Polarization dependent XPS measurements reveal substantial ligand orbital 
polarization.  An exchange splitting energy of $\Delta_{ex-spl}\simeq 3.2$ eV
was deduced~\cite{[5]}, implying substantial correlation effects. On the 
other hand, LSDA calculations yield $\Delta_{ex-spl}\simeq 1.8 eV$!

(2) The resistivity shows a behavior characteristic of correlated Fermi 
liquids~\cite{[6]}: $\rho(T)=\rho_{0}+AT^{2}+BT^{7/2}$, the last term coming 
from the carrier scattering of two-magnon fluctuations in a double-exchange
ferromagnet~\cite{[7]}.  A similar kind of behavior has also been observed in 
the CMR manganites in the FM metallic phase~\cite{[7]}.  In fact, $CrO_{2}$ 
falls into the ``bad metal'' classification, with high-$T$ 
($T>T_{c}^{FM}=390K$) resistivity exceeding the Mott limit~\cite{[10]}. The 
coefficient $A$ is large, and in fact the ratio $A/\gamma^{2}$ ($\gamma$ is 
the coeeficient of the linear term in the electronic specific heat) is close 
to the value expected for heavy-fermion metals~\cite{[8]}, implying
substantial correlation effects in the Cr $d$-band.

(3) Additional evidence comes from the optical conductivity data, which reveals
a Drude conductivity at low energies, followed by a broad bump around $0.8 eV$
and high-energy features centered around $3 eV$~\cite{[9]}. While the Drude 
part is understandable within the LSDA approach, the other features mentioned 
above are seen at energies that are systematically $10-20$ percent {\it lower}
than those predicted by the LSDA calculation, showing up the importance of 
correlation effects in this material~\cite{[9]}.

(4) Additionally, recent magnetotransport measurements show interesting 
features.  Firstly, the magnetoresistance is {\it linear} in fields upto $1$ 
Tesla,
and shows no evident relation with the magnetization~\cite{[10]}.  On the 
other hand, it was found that the longitudinal and transverse MR is negative 
and increases linearly with field for $T>200$K, while having a concave shape 
as a function of field for lower $T$~\cite{[10]}, passing through minima as 
functions of field around $T \simeq 200$K.  It is interesting to notice that 
similar behavior is also observed for manganites, $La_{1-x}Sr_{x}MnO_{3}$ 
with $x=0.175, 0.2$~\cite{[12]}. An understanding of the observed crossovers 
around $T=200$K requires the consideration of additional mechanism(s) over 
and above the double-exchange interaction, since these features occur deep 
inside the ferromagnetic metallic phase ($T_{c}=390$K).

(5) Existing photoemission data on $CrO_{2}$ show features that are more 
characteristic of a semiconductor with vanishing DOS at the Fermi level than 
of a good metal~\cite{[11]}, completely consistent with the "bad metal" 
classification made in the resistivity measurements. Such features, also 
observed in the metallic 
FM state of the CMR manganites, have been linked to the strong Jahn-Teller 
distortions known to exist for CMR manganites~\cite{[17]}. However, neither 
the static J-T distortion, nor the doping-induced static disorder is present 
in $CrO_{2}$, leaving one to look for alternative scenarios to understand 
the suppression of low-energy spectral weight in PES. 

  We start by noticing that the emergence of a coherent, strongly correlated
Fermi liquid scale required to understand the above features is out of reach
of LDA based calculations (within which a large Woods-Saxon ratio is
inexplicable, for example), as well as of calculations based on pure double
exchange models.  This is because these features are observed deep inside the
ferromagnetic metallic state, and are thus related to additional scattering 
mechanisms in a half-metallic situation (where one deals with a fully 
spin-polarized band).  In this paper, we argue that the above effects can 
indeed be understood by invoking the important role of strong local, orbital 
correlations in the $t_{2g}$ sector (see below).

  An understanding of features mentioned above should go hand-in-hand with the
basic electronic structure.  In this context, the LSDA+U calculation carried 
out by Korotin {\it et al.}~\cite{[3]} already provides the necessary 
ingredients to pursue more detailed investigations.  The necessary ingredients 
seem to be: (1) Strong, local Hubbard $U$ and Hund's rule coupling, $J_{H}$.
(2) Hybridization of the $d_{yz+xz}$ band with the $p$ band of the $O$, 
resulting in $Cr$ being mixed-valent, like $Mn$ in the manganites.
As pointed out in~\cite{[3]}, the fact that $CrO_{2}$ is in the class of 
negative charge-transfer gap materials makes this $d-p$ hybrid band 
self-doped.  The well known double-exchange mechanism then drives the FM 
metallic phase.

  An examination of Figs.~(3)-(4) of Ref.~\cite{[4]} shows that the the bands 
crossing the Fermi level have comparable $d_{xz+yz}$ and $d_{yz-zx}$ 
character, as well as a non-dispersive $d_{xy}$ band centered $1 eV$ below 
$\mu$.  We shall consider the $O$ band to act solely as a reservoir of 
carriers, serving to make the $d$-band manifold non-half-filled~\cite{[4]}.  
The almost complete spin polarization tells us that the local Hund's rule 
coupling is strong.  If this is so, we have to assume that the on-site Coulomb
repulsion (the intra-orbital Hubbard $U$) is stronger.  Given the above, we are
then led to consider a two-orbital Hubbard-like model in the double exchange
(DE) limit, $U,J_{H}>>t_{a,b}$ ($a,b$ label orbital indices).  The relevant 
minimal model for our purpose is written as,

\be
H=H_{1}+H_{2}
\ee
where

\be
H_{1}=-\sum_{<ij>\sigma}t_{ij}^{ab}(c_{ia\sigma}^{\dag} c_{jb\sigma}+h.c) +
U\sum_{i}(n_{ia\uparrow}n_{ia\downarrow} + a \rightarrow b)  
\ee
and

\be
H_{2}= U_{ab}\sum_{i\sigma\sigma'}n_{ia\sigma}n_{ib\sigma'} - 
J_{H}\sum_{i}{\bf S_{i}} \cdot ({\bf \sigma_{i}^{a}}+{\bf \sigma_{i}^{b}})
\ee

  In the above, we shall assume that the effects of the actual bandstructure 
have been incorporated into the effective hopping strengths by fitting the 
idealized bandstructure to the actual one; this has been employed with much
success for the cuprates~\cite{[13]}.  We shall henceforth treat the $t^{ab}$ 
as parameters of the effective model above.  $U_{ab}$ is the local, 
inter-orbital Coulomb repulsion, ${\bf S_{i}}$ is the localized spin in the 
$d_{xy}$ orbital, and that $a=d_{xz+yz}$ and $b=d_{yz-zx}$.  In the DE limit, 
the states  corresponding to a carrier aligned antiparallel to the core-spin 
are projected out, and so one can drop the spin indices in the above eqn, at 
the price of introducing a magnetization-dependent hopping, 
$t_{ij}^{ab}=t_{ij}^{ab}({\bf S})$.  In this situation, the orbital index 
plays the role of the spin, and so one is finally left with a generalized 
Hubbard model in orbital space.  To proceed, we will further make an 
assumption which does not qualitatively affect the final conclusions (see 
below).  We will assume that the hopping matrix is diagonal in orbital space, 
$t_{ij}^{ab}=t_{ij}\delta_{ab}$~\cite{[14]}, and choose
$t^{aa}=t_{1}$ and $t^{bb}=t_{2}$.  Finally, we introduce new fermion 
operators, $c_{\uparrow}=(c_{a}+c_{b})/\sqrt{2}$, and
$c_{\downarrow}=(c_{a}-c_{b})/\sqrt{2}$.  The Hamiltonian now becomes,

\be
H=-\sum_{<ij>\sigma}t_{\sigma}({\bf S})(c_{i\sigma}^{\dag} c_{j\sigma}+h.c) +
 U\sum_{i}n_{i\uparrow}n_{i\downarrow}                    
\ee
At low-$T$ ($<T_{c}^{FM}=390$K), the hopping is a function of the 
magnetization, 
$t_{\sigma}({\bf S})=t_{\sigma}(M)=
t_{\sigma}[1+<S_{i}^{z}S_{j}^{z}>/2S^{2}]^{1/2}$ 
with $M=N^{-1}\sum_{i}<S_{i}^{z}>$, so one has to deal with a Hubbard model 
in orbital space with an orbital pseudo-spin and magnetization 
dependent hopping. In the above eqn.~(4), $t_{\uparrow}=(t_{1}+t_{2})$, and 
$t_{\downarrow}=(t_{1}-t_{2})$.  The case $t_{1}=t_{2}$ has been considered 
in~\cite{[14]}, where the anomalous FM metallic phase of the manganites is 
sought to be understood in terms of the non-FL state of the effective 
Falicov-Kimball model in $d=\infty$~\cite{[15]}.  An exact solution of the 
FKM in this limit reveals that the non-FL behavior is driven by a divergence 
in the local excitonic correlation function at low energy. In contrast, there 
is evidence that the FM metallic phase in $CrO_{2}$ as well as in 
$La_{0.7}Sr_{0.3}MnO_{3}$ is a correlated Fermi liquid at low $T$, requiring 
us to work with the model, eqn.~(4).

In what follows, we will explain much of the observed features mentioned above
in terms of the $d=\infty$ solution of the effective Hubbard model (eqn.~(4)).
To understand where the local FL behavior comes from, we notice that the model,
eqn.~(4) with $t_{1}=t_{2}$, can be mapped onto the x-ray edge problem with 
the recoilless local scatterer (the $\downarrow$-spins) right at the Fermi 
surface.  In this case, the spectral fn. of the $\downarrow$-spin fermion, as 
well as that of the ``exciton''~\cite{[15]} is divergent at low energy. The 
case $t_{1} \ne t_{2}$ corresponds to allowing the scatterer to recoil, 
killing the infrared divergence and restoring FL behavior at low $T$ 
($T<E_{R}$, the recoil energy of the $\downarrow$-spin~\cite{[15]}).  In the 
$d=\infty$ solution of the model, however, FL coherence sets in below the 
lattice coherence scale, $T_{coh}$ (related to $t_{1}-t_{2}$). 

  It is instructive to summarize what is known about the Hubbard-like model, 
eqn.~(4), in the $d=\infty$ limit~\cite{[15]}.  Above $T_{coh}$, the metallic 
state off half-filling is not a Fermi liquid, and the dynamics of the carriers 
is determined by inelastic scattering off the effectively unquenched local 
moments (orbital moments in our approach).  Below $T_{coh}$, the local moments 
are screened by the ``conduction electron'' spin density, giving a Fermi liquid
response.  This coherence scale can be driven quite low for given $t_{1}-t_{2}$
resulting in an anomalous response at not too low $T$.  In particular, the
suppression of quasiparticle features in the local DOS for 
$T>T_{coh}$~\cite{[16]} would result in: (a) a non-Drude optical response, 
(b) a pseudogap feature in photoemission, (c) absence of the $T^{2}$ term in 
the resistivity and an anomalous $T$-dependence of the Hall 
constant~\cite{[16]}, and (d) lastly, one would expect the FL coherent 
response below $T_{coh}$ to manifest itself in these responses as $T$ is 
lowered.  In particular, below $T_{coh}$, one expects a $T^{2}$ component in 
the resistivity, a low energy Drude response in the optical conductivity, a 
coherent feature in the photoemission lineshape at the Fermi surface, and a 
$T$-independent normal Hall constant (the anomalous Hall effect requires more 
work~\cite{[18]}).

  Given that the para-orbital metallic state of the above model with 
$t_{1} \ne t_{2}$ is a correlated Fermi liquid, one expects perturbations 
like disorder to tilt the balance in favor of low-energy incoherence.  In 
particular, with modest disorder~\cite{[19]}, the incoherent metallic state 
gives rise to an incoherent lineshape in PES and a pseudogapped behavior in 
the optical response, even at low-$T$.

  With this information in mind, we are ready to discuss our results.  In our
calculations, we have employed a gaussian unperturbed DOS, and two values of  
temperature: $T=0.01D$ and $T=0.1D$ ($D$ is the half-width of the gaussian).
We have chosen $U_{ab}=3D_{\uparrow}$, where $D_{\uparrow}$ is the effective 
$\uparrow$-spin bandwidth, and the bandfilling, 
$n=n_{\uparrow}+n_{\downarrow}=0.8$.  In the $d=\infty$ approximation that we 
use, $t_{\sigma}(M)=t_{\sigma}[(1+M^{2}(T))/2]^{1/2}$ with $M(T)$ taken from 
the spherical model limit for the Heisenberg model.

In $d=\infty$, all relevant information about the local dynamical fluctuations
is encoded in a purely local irreducible self-energy, $\Sigma(\omega)$, 
entering the lattice Green function for the model under consideration:
\be
G(k,\omega)=G(\epsilon_{k},\omega)=\frac{1}{\omega+\mu-\epsilon_{k}
-\Sigma(\omega)}
\ee
To solve the model in $d=\infty$ requires a reliable way to solve the single 
impurity Anderson model (SIAM) embedded in a dynamical bath described by the
hybridization fn. $\Delta(\omega)$.  There is an additional condition that 
completes the selfconsistency:
\be
\int d\epsilon G(\epsilon,\omega)\rho_{0}(\epsilon) = 
\frac{1}{\omega+\mu-\Delta(\omega)-\Sigma(\omega)}
\ee
where $\rho_{0}(\epsilon)$ is the free DOS ($U=0$).  The above eqns.~(4)-(6) 
refer to the disorder-free Hubbard model.  In $d=\infty$, this is sufficient 
to compute the transport, because the vertex corrections in the Bethe Salpeter 
eqn. for the conductivity vanish identically in this limit~\cite{[15]}. Thus,
the conductivity is fully determined by the basic bubble diagram made up of
fully interating local GFs of the lattice model.

The optical conductivity is computable in terms of the full $d=\infty$ GFs as 
follows~\cite{[15]}:
\be
\sigma_{xx}(i\omega)=\frac{1}{i\omega}\int 
d\epsilon\rho_{0}(\epsilon)\sum_{i\nu}G(\epsilon,i\nu)G(\epsilon,i\nu+i\omega)
\ee
and the dc resistivity is just $\rho_{dc}(T)=1/\sigma_{xx}(0,T)$.  

To solve the impurity model, we have used the iterated perturbation theory 
adapted to our model.  We use the IPT because it is analytic in $U$ and 
yields results in good agreement with exact diagonalization 
studies~\cite{[15]}.
In view of the ability of the IPT to reproduce all the qualitative aspects 
observed in $\sigma_{xx}(\omega)$, we believe that is a good tool 
in the present case. We mention that we have extended the IPT off half-filling 
to finite temperatures (as far as we are aware, such an undertaking has not 
been carried out earlier).  We have checked that our results agree fully with 
those obtained by others at $T=0$.

The photoemission lineshape is given simply by,
\be
I_{PES}(\omega)=\rho(\omega-\mu,T)f(\omega-\mu)
\ee
where $f(\omega)=(e^{\omega/T}+1)^{-1}$ is the Fermi fn. 

To make closer contact with experiment, we notice that, in 
$CrO_{2}$, $Cr$ exists in two crystallographically inequivalent sites, 
resulting from a particular kind of orbital ordering~\cite{[3]}. The 
potential fluctuation arising from this is of order of the crystal field 
splitting ($\simeq 1-1.5$ eV), providing an intrinsic source of disorder 
scattering (especially at $T\simeq 300$K at which the PES experiment was 
performed), further reducing the coherent spectral weight at $\mu$.
Hence, we have considered the effects of scattering off an additional local, 
static and random potential $v=U/4$, assuming a binary distribution, 
$P(v_{i})=[{\delta(v_{i})+\delta(v_{i}-v)}]/2$ (an additional assumption is 
that the material is composed of an equal composition of the two inequivalent 
sites).  To this purpose we have used the IPT (for correlations) in 
combination with the CPA for disorder, in a selfconsistent way~\cite{[19]}. 

\begin{figure}[htb]
\epsfxsize=3.3in
%%\epsffile{cro1.eps}
\epsffile{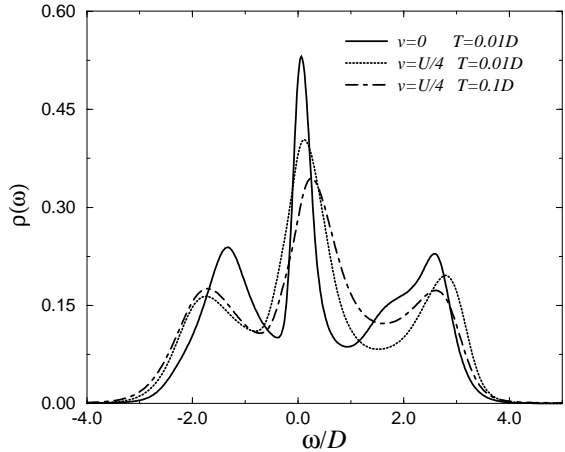}
\caption{The density of states (DOS) for $U/D=3.0$, $\delta=0.2$, $T=0.01D$ 
and two values of local disorder potential: $v=0$ (continuous) and 
$v=U/4$ (dotted line).  The dot-dashed line is the DOS for the same 
parameters as above and $v=U/4$ but with $T=0.1D$.}
\label{fig1}
\end{figure}

In Fig.~\ref{fig1}, we show the local spectral density,
$\rho(\omega)$ for our model at different $T$ for $U/D=3.0$  This ratio was
chosen because we are interested in dealing with the metallic phase here.
It is clearly seen that the collective ``Kondo'' resonance is strongly $T$
dependent because of particle-hole pair generation; above 
$T\simeq T_{coh}=0.13$, the coherent feature is completely washed out, and 
the physics is understood in terms of carriers scattered inelastically off 
effectively unquenched {\it orbital} moments, in analogy with the situation 
in the usual Hubbard model. This has important consequences for the optical 
and spectroscopic response, as we describe below.

\begin{figure}[h]
\epsfxsize=3.3in
\epsffile{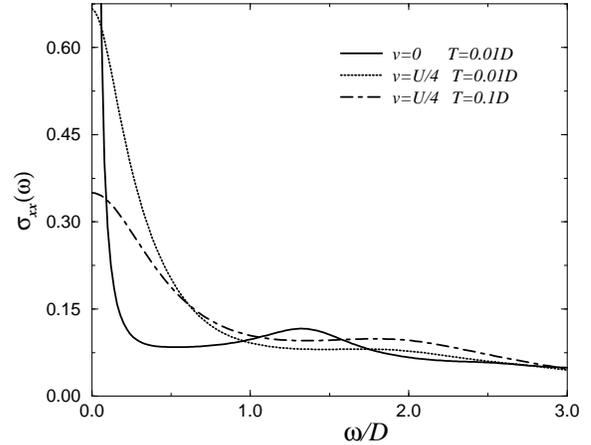}
\caption{Real part of the optical conductivity 
$\sigma_{xx}$ for $U/D=3.0$, $\delta=0.2$ $v=0$ (continuous) 
and $v=U/4$ (dotted), all at $T=0.01D$. The dot-dashed line shows
$\sigma_{xx}$ for $U/D=3.0$, $\delta=0.2$, $v=U/4$ and $T=0.1D$.
The computed result shows all the features observed in Ref.~[9].}    
\label{fig2}
\end{figure}

  At low $T$, the optical conductivity shows a sharp Drude peak characteristic
of a correlated FL metal.  This peak carries a small part of the total spectral
weight, and most of it is distributed over a wide energy scale of $3 eV$,
comprising the rather well pronounced mid-IR absorption and higher energy
features (Fig.~\ref{fig2}). Increasing $T$ ($T=0.1D$), we see that the Drude 
weight is reduced and the higher energy features are smoothened out, again 
resembling qualitatively the situation in $CrO_{2}$~\cite{[9]}.

  The d.c resistivity (not shown) 
is computed from the $\omega=0$ part of the optical 
conductivity, and clearly exhibits the quadratic behavior in $T$:
\be
\rho_{dc}(T) \simeq \tau^{-1}(T)=\pi^{2}(U^{2}/D^{3})n/2(1-n/2)T^{2}
\ee
  With an effective mass enhancement of approximately $3$ for $U_{ab}/D=3.0$,
$n=1.3.10^{28}m^{-3}$, $\rho(\mu)=0.55 eV^{-1}$ and $n=0.8$, we obtain 
$A/\gamma^{2} \simeq 4-5*10^{-5}$, as reported in Ref.~\cite{[8]}.
   
\begin{figure}[h]
\epsfxsize=3.3in
\epsffile{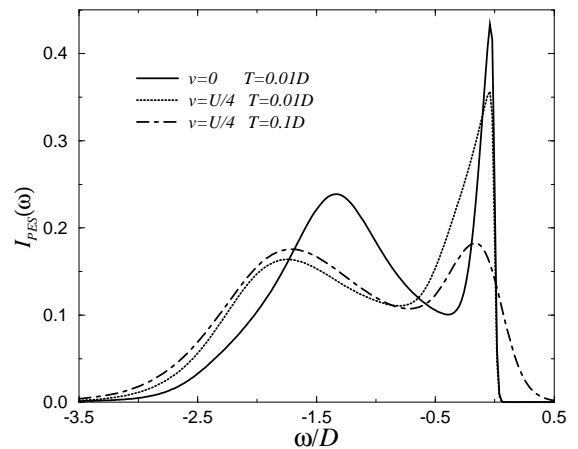}
\caption{The integrated photoemission lineshapes for our model with $U/D=3.0$, 
$\delta=0.2$ and for two values of the disorder strength $v$ and the 
temperature $T$. Introduction of disorder brings the lineshape much closer to 
that observed experimentally by moving low energy spectral weight to higher 
energy.} 
\label{fig3}
\end{figure}

  Next we turn to angle integrated photoemission, which measures the occupied 
part of the single-particle DOS, in Fig.~\ref{fig3}. As expected from the 
$T$-variation of the DOS, we 
see a sharp Fermi step at $T=0.01D$, a reduced step feature at 
$T=0.1D$, and completely incoherent response in the pseudogap phase.  A
comparison of the results with those observed experimentally shows that the 
calculated  spectral weight at the Fermi surface ($\omega=0$) is quite 
\hspace{-.4cm}small,
but still finite, while {\it complete} depletion of the low energy spectral 
weight is observed experimentally.  The almost complete depletion of the 
low-energy spectral weight observed experimentally is then interpreted in 
terms of the $T$-dependence of the local spectral density.  If the temperature 
of measurement is higher than the lattice coherence scale, the FL-like feature 
is washed out, reducing the spectral weight at $\mu$, and making the intensity 
resemble that of a semiconductor.  In this context, we would like to mention 
that the PES experiment was carried out at $T=300$ K, which is pretty high.
 The calculation carried out above does yield an incoherent response at 
$T=0.1D$ and the spectral weight at $\mu$ is quite small
but is still non-vanishing at $\mu$.

As we have mentioned before, the results for a moderate disorder $v=U/4$ 
resemble experimental observations more closely (fig.~3).
However, it may also be the case that short-ranged orbital correlations 
enhance the tendency to open up a pseudogap, reducing the DOS beyond that 
found in the $d=\infty$ calculation. At this moment, this discrepancy between 
theory and experiment remains unclear. A more precise undertaking should also 
include the real bandstructure of $CrO_{2}$; in this context, we notice that 
the LDA calculations already yield a dip in the DOS at $E_{F}$~\cite{[3]}. We 
are of the opinion that the real situation involves a combination of all the 
above factors, and plan to address these in a future publication. We suggest 
that more information concerning the $T$-dependence of the low-energy spectral 
weight, could be obtained from a simultaneous study of optical and PES spectra 
taken at different temperatures across the ferro-para transition. This could 
be used to discriminate between different scenarios proposed for $CrO_{2}$.
If the ideas proposed here are correct, one expects a $T$-dependent build-up 
of coherent spectral weight at low energy as $T$ is lowered through the 
para-ferro transition.

  Finally, we observe that the model hamiltonian proposed in this paper can 
explain the orbital polarization observed by Stagarescu
{\it et al.}~\cite{[5]}. Indeed, for a generic choice of the 
$t_{\alpha}, \alpha=1,2$, with $t_{1}\ne t_{2}$, an orbital polarization is 
spontaneously generated in our model.  A complete analysis of these effects 
requires consideration of actual bandstructure within the $d=\infty$ 
approximation that we've used here, and we plan to treat this issue in the 
future.

  In conclusion, we have shown, based on the LDA+U results of Korotin 
{\it et al.}~\cite{[3]}, how the various observed ``strongly correlated'' 
properties of the half-metallic ferromagnet $CrO_{2}$ can be understood by 
invoking the role of dynamical orbital correlations and Hund's rule double 
exchange in the $t_{2g}$ sector.  The treatment presented here should also be 
valid for other transition metal based half-metallic ferromagnets~\cite{[2]}.
Our results are consistent with the view that LSDA calculations~\cite{[3]}
need to be supplemented by treatments including dynamical correlation effects 
to understand the physics of $CrO_{2}$. 

\acknowledgments
MSL acknowledges the financial support of the SfB341 of the German DPG.
LC was supported by the Funda\c c\~ao de Amparo 
\`a Pesquisa do Estado de S\~ao Paulo (FAPESP).


\begin{references}

\bibitem{[1]} for a review, see, for example, {\it Colossal Magnetoresistive 
Oxides}, ed. Y. Tokura (Gordon and Breach) in press. 

\bibitem{[2]} M. Imada, cond-mat/0004232.

\bibitem{[3]} M. Korotin {\it et al.}, Phys. Rev. Lett. {\bf 80}, 4305 (1998). 
I also thank Prof. Sawatzky for correspondence related to the additional 
scattering arising from a particular kind of orbital ordering found in their
calculation.

\bibitem{[4]} I. Mazin, D. J. Singh and C. Ambrosch-Draxl, Phys. Rev. B 
{\bf 59}, 411 (1999). 

\bibitem{[5]} A. Stagarescu, {\it et al.}, cond-mat 9910346, to be published 
in Phys. Rev. B. 

\bibitem{[6]} L. Ranno, A. Barry and J. M. D. Coey, J. Appl. Phys. {\bf 81},
 (8), 5774 (1997). 

\bibitem{[7]} V. Yu. Irkhin and M. Katsnelson, Phys. Usp. {\bf 37}, 659 (1994).
 
\bibitem{[8]} K. Suzuki and P. Tedrow, Phys. Rev. B{\bf 58}, 17, 11597 (1998).

\bibitem{[9]} D. Basov, {\it et al.}, Phys. Rev. B{\bf 60}, 4126 (1999). 

\bibitem{[10]} see ref.[8]; also A. Barry, Ph.D thesis, University of Dublin,
unpublished. 

\bibitem{[11]} K. K\"amper {\it et al.}, Phys. Rev. Lett. {\bf 59}, 2788 
(1987).

\bibitem{[12]} A. Ramirez, J. Phys. Condens. Matter. {\bf 9}, 8171 (1997).

\bibitem{[13]} see E. Stechel, in {\it High Temperature Superconductivity}, 
eds. B. Coffey {\it et al.}, (Addison-Wesley, 1989).

\bibitem{[14]} the actual situation is a bit more complicated than what we 
have assumed here, but in the ferromagnetic metallic state, we expect the 
qualitative features to remain unchanged.  See, for example, V. Ferrari, 
{\it et al.}, cond-mat/9906131.

\bibitem{[15]} A. Georges {\it et al.}, Revs. Mod. Phys.{\bf 68}, 13 (1996).

\bibitem{[16]} P. Majumdar and H. R. Krishnamurthy, cond-mat 9512151.

\bibitem{[17]} see, for example, A. J. Millis, P. B. Littlewood and B. I. 
Shraiman, Phys. Rev. Lett. {\bf 74}, 5144 (1995).

\bibitem{[18]} F. E. Maranzana, Phys. Rev. {\bf 160}, 421 (1967). 

\bibitem{[19]} M. S. Laad, L. Craco and E. M\"uller-Hartmann, cond-mat/9911378,
submitted to Phys. Rev. B.


\end{references}
\end{document}